\documentstyle[mprocl]{article}

\def\be{\begin{equation}}
\def\ee{\end{equation}}
\def\bea{\begin{eqnarray}}
\def\eea{\end{eqnarray}}

\begin{document}

\title{
MULTIDIMENSIONAL 
GRAVITY AND COSMOLOGY: EXACT SOLUTIONS} 

\author{ V.N. MELNIKOV }

\address{
Centre for Gravitation and Fundamental Metrology, VNIIMS,
\\ 3-1 M. Ulyanovoy Str., Moscow, 117313, Russia}

\maketitle\abstracts{}


\section{Introduction}

It is well known that pioneering papers of T.K. Kaluza and O. Klein
\cite{Kal,Kl} on 5-dimensional gravity 
(see also \cite{DeS,Lee,Vl1,WeP1})
opened the interest to 
investigations in multidimensional gravity. These ideas were continued by 
P.Jordan \cite{J} who suggested to consider the more general case 
$g_{55}\ne{\rm const}$ leading to the theory with an additional scalar 
field. The papers \cite{Kal,Kl,J} were in some sense a source of 
inspiration for C.Brans and R.H.Dicke in their well-known work on 
the scalar-tensor gravitational theory \cite{BD}. 

In the 70th an interest to multidimensional gravitational models was
stimulated mainly by: i) the ideas of gauge theories leading to the
non-Abelian generalization of Kaluza-Klein approach \cite{SaSe} and by ii)
supergravitational theories 
\cite{CJS,SaSe}. In the 80th the supergravitational
theories were "replaced" by superstring models \cite{GrSW}. In all
these theories 4-dimensional gravitational models with extra fields
were obtained from some multidimensional model by dimensional reduction
based on the decomposition of the manifold
$$M=M^4\times M_{int},$$
where $M^4$ is a 4-dimensional manifold and $M_{int}$ is some internal
manifold (mostly considered as a compact one). 

The earlier papers on multidimensional cosmology dealt with a
block-diagonal cosmological metric defined on the manifold
$$
M=R\times M_0\times ...\times M_n
$$
of the form 
$$
g=-dt\otimes dt+\sum_{r=0}^n a_r^2(t)  g^r
$$
where $(M_r,g^r)$ are Einstein spaces, $r=0,\dots,n$ \cite{FH}
-\cite{Mel2}.

In \cite{GRT,BL1,BL3,BIM1,Mel2,IM2,BLP,IM5,GIM} the models with higher 
dimensional "perfect"-fluid were considered. In these models pressures 
(for any component)
are proportional to the density 
$$ p_r=\left(1-\frac{u_r}{d_r}\right)\rho, 
$$ 
$r=0,\dots,n$, where $d_r$ is a dimension of $M_r$. Such models are 
reduced to pseudo-Euclidean Toda-like systems with the Lagrangian 
$$ 
L=\frac12G_{ij}\dot x^i\dot x^j-\sum_{k=1}^mA_k{\rm e}^{u_i^kx^i}
$$
and the zero-energy constraint $E=0$. 
In a classical case  exact solutions 
with Ricci-flat $(M_r,g^r)$ for 1-component case were considered by many 
authors 
(see, for example, \cite{Lor1,BO,BIM1,IM2,IM5,GIM} and references 
therein).  For the two component perfect-fluid there were solutions with 
two curvatures, i.e.  $n=2$, when $(d_1,d_2)=(2,8),(3,6),(5,5)$ 
\cite{GIM2} 
and corresponding non-singular solutions from \cite{IM14}. Among the 
solutions \cite{GIM2} there exists a special class of Milne-type 
solutions.  Recently some interesting extension of  
2-component solutions were obtained in \cite{GI}.

It should be noted that the pseudo-Euclidean Toda-like systems are not 
well-studied yet. There exists a special class of eqs. of state that gives 
rise to the Euclidean Toda models. First such solution was considered in 
\cite{GIM} for the Lie algebra $a_2$. Recently the case of $a_n=sl(n+1)$ 
Lie algebras was considered and the solutions were expressed in terms of 
a new elegant representation obtained by Anderson \cite{GKMR}.

The cosmological solutions may have regimes with: i) spontaneous and 
dynamical compactifications; ii) Kasner-like behavior near the 
singularity; iii) izotropization for large times (see, for example, 
\cite{BIMZ,IM5}).

Near the singularity one can have an oscillating behavior like in 
the well-known mixmaster  (Bianchi-IX) model. Multidimensional 
generalizations of this model were considered by many authors (see, for 
example, \cite{BK,BS,DHS,SzP}). In \cite{IKM,IM6} the billiard 
representation for multidimensional cosmological models near the 
singularity was considered and the criterion for the volume of the 
billiard to be finite was established in terms of illumination of the unit 
 sphere by point-like sources. For perfect-fluid this was considered in 
detail in \cite{IM6}. Some interesting topics related to general 
(non-cosmological) situation were considered in \cite{KM}.

Multidimensional cosmological models have a generalization to the 
case when the viscosity of the "fluid" is taken into account 
\cite{GMT}. Some  classes of exact solutions were 
obtained, in particular nonsingular cosmological solutions.

Multidimensional quantum cosmology based on the Wheeler-DeWitt (WDW) 
equation
$$
\hat H\Psi=0,
$$
where $\Psi$ is the so-called "wave function of the universe", was 
treated first in \cite{IMZ}, (see also \cite{Hal}). This equation was 
considered for the vacuum case in \cite{IMZ} and integrated in a very 
special situation of 2-spaces. The WDW equation for 
the "perfect-fluid" was investigated
in \cite{IM5}. The exact solutions in 1-component case were 
considered very carefully in \cite{IM10} (for perfect fluid). In 
\cite{Zh2} the multidimensional quantum wormholes were 
suggested, i.e.  solutions with a special-type behavior  
of the wave function (see \cite{HawP}).

These solutions were generalized to the perfect-fluid case in 
\cite{IM5,IM10}.  In \cite{IM6} the "quantum billiard" was obtained 
for WDW solutions near the singularity. 
It should be also noted that the "third-quantized" 
multidimensional cosmological models were considered in several papers
\cite{Zh1,Hor,IM10}.

Cosmological solutions are 
closely related to solutions with the spherical symmetry. The first 
multidimensional generalization of such type was considered by D. Kramer 
\cite{Kr} and rediscovered by A.I. Legkii \cite{Le},
D.J.Gross and M.J.Perry  \cite{GP} ( and also by Davidson and Owen). 
In \cite{BrI} the Schwarzschild solution was generalized to 
the case of $n$ internal Ricci-flat spaces and it was shown that black 
hole configuration takes place when scale factors of internal spaces are 
constants. In \cite{FIM2} an analogous generalization of the Tangherlini 
solution \cite{Tan}
was obtained. These solutions were also generalized to the 
electrovacuum case \cite{FIM3,IM8,BM}. In \cite{BI,BM}
multidimensional dilatonic black holes were singled out.
An interesting theorem was proved in \cite{BM}  
that "cuts" all non-black-hole configurations as non-stable. In 
\cite{IM13} the extremely-charged dilatonic black holes were generalized 
to Majumdar-Papapetrou case when the cosmological constant is non-zero.

At present there exists a special interest to 
the so-called M-,F-theories etc.
\cite{Sc,Du}.
These theories are "supermembrane" analogues of superstring models in
$D=11,12$ etc. \cite{GrSW}. 
The low-energy limit of these theories leads to models
governed by the action
\begin{equation}
\label{1.2}
S=\int d^D x \sqrt{\mid g\mid} 
\{ {R}[g] - h_{\alpha\beta}\;
g^{MN} \partial_{M} \varphi^\alpha \partial_{N} \varphi^\beta
\\ 
 - \sum_{a \in \Delta}
\frac{\theta_a}{n_a!} \exp[ 2 \lambda_{a} (\varphi) ] (F^a)^2_g \},
\end{equation}
where $g$ is metric, $F^a=dA^a$ are forms of rank $F^a=n_a$, and 
$\varphi^\alpha$ are scalar fields.

In \cite{IM16} it was shown that after dimensional reduction on the 
manifold $M_0\times M_1\times\dots\times M_n$ and when the
composite $p$-brane ansatz is considered the problem is reduced to the 
gravitating self-interacting $\sigma$-model with certain constraints 
imposed. For electric $p$-branes see also \cite{IM11,IM12,IMR}.
This 
representation may be considered as a powerful tool for obtaining 
different solutions with intersecting $p$-branes (analogs of membranes). 
In \cite{IM16,IMR} the Majumdar-Papapetrou type solutions were obtained 
(for non-composite case see \cite{IM11,IM12}). These solutions correspond 
to Ricci-flat $(M_i,g^i)$, $i=1,\dots,n$, and were generalized also to the 
case of Einstein internal spaces. Earlier some special classes of these 
solutions were considered in 
\cite{Ts1,GKT,AR,AEH,AIR}. The obtained solutions take 
place, when certain orthogonality relations (on couplings parameters, 
dimensions of "branes", total dimension) are imposed. In this 
situation a class of cosmological and spherically-symmetric solutions was 
also obtained \cite{IM18}. Special cases were also considered in 
\cite{LPX,BGIM,BKR}. The solutions with the horizon were also 
considered in details in \cite{CT,AIV,Oh}. In \cite{BIM3} some 
propositions related to i) interconnection between the Hawking temperature 
and the singularity behaviour, and ii) to multitemporal configurations 
were proved.

It should be noted that the multidimensional and multitemporal
generalization of the 
Schwarz\-schild and Tangherlini solutions were considered in 
\cite{IM8,IM9}, where the generalization of Newton's formulas on 
multitemporal case was obtained.

We note also that  there exists a large variety of Toda solutions 
(open or closed) when certain intersection rules are satisfied \cite{IM18}.

In \cite{IM18} the Wheeler-DeWitt equation was also integrated for 
intersecting $p$-branes (in orthogonal case). The slightly-different 
approach was considered also in \cite{LMMP}. These solutions may be 
considered as an interesting first step for quantum description of 
low-dimensional supergravity theories in different super-$p$-branes 
theories.

Problems of multidimensional gravitational theories and models
were discussed at a variety of conferences during last years such as 
Marcel Grossmann meetings, conferences in Russia in Jaroslavle (1994),
Novgorod (1996) and Ulyanovsk (1997). The whole session was devoted to 
these topics at the 8 Marcel Grossmann meeting in Jerusalem (June 1997).

\section{Review of reports}

Here are  some short review of main talks submitted and discussed 
at this section.

The talk of E.I.Guendelman and A.B.Kaganovich was devoted to
"Gravitational Theory without the Cosmological Constant Problem".
It is known that 
the vacuum energy density or cosmological constant is predicted by modern
particle theory to be very large \cite{W}. Such parameter controls the
large scale structure of the universe, which is consistent with a zero
cosmological constant. A new approach for the solution
of the cosmological constant problem developing a new
gravitational theory \cite{GK1} - \cite{GK3}
is offered  where the measure
of integration in the action principle is not necessarily $\sqrt{-g}$
but it is determined dynamically through additional degrees of freedom. This
theory is based on the demand that such measure respect  the principle of
"non gravitating vacuum energy" \cite{GK1}  which states that the
Lagrangian density $L$ can be changed to $L+constant$ without affecting
the dynamics. Formulating the theory using the metric and the affine
connection, as well as the fields
which define the measure of integration, as independent dynamical
variables  they get as a consequence of the
variational principle a constraint that enforces the vanishing of the
cosmological constant \cite{GK2}, \cite{GK3}. A successful
model \cite{GK3} that implements these ideas is realized in a
six or higher
dimensional space-time. The compactification of extra dimensions into a
sphere gives the possibility of generating scalar masses and potentials,
gauge fields and fermionic masses. It turns out that remaining four
dimensional space-time must have effective zero cosmological constant.
Another model \cite{GK4}, based on the same principles, but
formulated in
four-dimensions, can incorporate a period of inflation for the early
universe before making a transition to a zero cosmological constant phase
which is realized without fine tuning.

The talk of V. D. Ivashchuk and V. N. Melnikov,
"Exact Solutions in Multidimensional Gravity with
Intersecting p-branes"  was devoted to
multidimensional gravitational models
containing  several dilatonic scalar fields
and antisymmetric forms (see (\ref{1.2})). The manifold
is chosen in the form $M = M_0 \times M_1 \times \ldots
\times M_n$,
where  $M_i$ are Einstein spaces ($i \geq 1$).
The block-diagonal metric is chosen and all fields and scale 
factors of the metric are functions on $M_0$.
For the forms composite (electro-magnetic) p-brane ansatz is adopted.
The model is reduced to gravitating self-interacting sigma-model
with certain constraints.
In pure electric and magnetic
cases the number of these constraints is $n_1(n_1 - 1)/2$
where $n_1$ is the number of 1-dimensional manifolds among $M_i$.
In the "electro-magnetic" case for ${\rm dim} M_0 = 1, 3$ additional
$n_1$ constraints appear.
A family of "Majumdar-Papapetrou type"  solutions
governed by a set of harmonic functions
is obtained, when all factor-spaces $M_{\nu}$ are Ricci-flat.
These solutions are generalized to the case of
non-Ricci-flat $M_0$ when also some additional
"internal" Einstein spaces of non-zero curvature are
added to $M$.  As an example
exact solutions for  $D = 11$ supergravity and
related 12-dimensional theory \cite{KKP} are presented.

In a paper of S.S.Kokarev  "New point of view on space-time dynamics"
it was suggested, that a curved 4-dimensional
space-time manifold is a surface of a strained
elastic plate in multidimensional embedding space-time. All thicknesses
$h_{m}$ of the plate along extra dimensions are much less
than its 4-dimensional sizes.
Multidimensional elastic free energy density integrated
over extra (internal) coordinates
of the strained plate has the form
\begin{equation}\label{freegen}
F_{\rm pl}=\frac{\mu H_{N}h_{m}^{2}}{12}\eta^{mn}\int \sqrt{-g}d^{4}x
\left\{\xi_{m,\mu,\nu}\xi_{n,\ ,}^{\ \mu\ \nu}+
f\xi_{m,\mu,}^{\ \ \mu}\xi_{n,\nu,}^{\ \
\nu}\right\}.
\end{equation}
where $\mu, f$ are "phenomenological" elastic constants of the theory,
$H_{N}$ is the  product of all thicknesses of the plate,
$\xi^{m}$ are components of a displacement vector,
$\eta_{mn}$ is the part of a metric of the embedding space,
$g_{\mu\nu}$ is a metric on the  plate surface.
This expression has been derived for a weak straining case,
and is based on the multidimensional Hook's law.
It is similar to a  Lagrangian density of a gravitational
field in GR up to a multiplicative constant,
if the thicknesses  in extra dimensions
are equal to each other,
the Poisson coefficient
of a material of the space-time plate
$\sigma$
is equal to 1/2 $(f=-1)$.
Dimensional analysis gives for the Yung modulus of a space-time 
substance value  $\sim10^{10^2}$Pa.  Variation of the (\ref{freegen}) over 
$\xi^{m}$ leads to  bewave equation for displacement vector components $$ 
D_{m}\Box^{2}\xi^{m}=P^{m}, \ \  \mbox{(no summation!)}
$$
and to the boundary conditions at the boundary of the plate:
$$
D_{m}\oint_{\partial\Sigma}d^{3}S^{\mu}\Box\xi_{m,\mu}\delta\xi^{m}+
\frac{D_{m}}{f+1}
\oint_{\partial\Sigma}d^{3}S^{\mu}(\xi_{m,\mu,\lambda}+
\eta_{\mu\lambda}f\Box\xi_{m})\delta\xi^{m\ \lambda}_{,}=0.
$$
Here
$D_{m}$ is the cylindrical stiffness factor of the 4-dimensional plate
in $m$-th extra dimension $(m=\overline{1,N})$,
depending on elastic constants of the theory,
$N$ is the number of additional to
classical four dimensions of the embedding space,
$P_{m}$ is normal to $V_{4}$ component of an external multidimensional
forces, bending the plate.

GR in such approach corresponds to degenerate case,
since under $\sigma=1/2$
(and only in this case!)
free energy density is a total divergence.
Basic notions and relations of GR acquire in this approach 
a new interpretation.
For example, Einstein equations appear as some special relation between
normal and tangent  stresses of the space-time plate.

The talk by U.G\"unther and A.Zhuk
"Gravitational Excitons From Extra Dimensions"
was devoted to stability conditions for compactified
internal spaces. Starting from a multidimensional cosmological model 
after a dimensional reduction  an effective four-dimensional
theory in Brans - Dicke and Einstein frames was obtained. 
The Einstein frame was
considered  as the physical one. In this frame they derived an
effective potential. It was shown that small excitations of the scale
factors $a_i = \exp \left( \beta^i \right) , \quad (i=1,\ldots ,n) $
of internal spaces near minima of the effective potential have a
form of massive scalar particles (gravitational excitons) developing in the
external space - time. The exciton masses strongly depend on the dimensions
and curvatures of the internal spaces, and possibly present additional
fields living on the internal spaces. These fields will contribute to the
effective potential, e.g. due to the Casimir effect, and by this way affect
the dynamics of the scale factor excitations. So, the detection of the scale
factor excitations can not only prove the existence of extra dimensions, but
also give additional information about the dimension of the internal spaces
and about fields possibly living on them.

For some particular classes of effective potentials with one, two and $n$
scale factors exciton masses as functions of parameters of the
internal spaces were calculated 
and  stability criterions necessary for the
compactification of the spaces were derived.

Their analysis shows that conditions for the existence of stable
configurations may depend not only on dimension and topology of the internal
spaces, and additional fields contributing to the effective potential, but
also on the number of independently oscillating scale factors. For example, $%
n-$scale-factor models with a saddle point as extremum of the effective
potential $U_{eff}(\beta _1,\ldots ,\beta _n)$ would lead to an unstable
configuration. Masses of the corresponding excitations would be positive
(excitons) as well as negative (tachyons). Under scale factor reduction to
an $m-$scale factor model with $m<n$, i.e. when  some of the scale
factors  were connected
by constraints $\beta _i=\beta _k$, the saddle point may, for
certain potentials, reduce to a stable minimum point of the new effective
potential $U_{eff}(\beta _1,\ldots ,\beta _m)$. As a result all masses of
excitations would be positive (excitons). This
''stabilization via scale factor reduction'' 
is demonstrated explicitly on a model with
one-component perfect fluid.

In Paulo Gali Macedo talk
"Anti-gravity Effect in Jordan-Thiry Unified Field Theory"
it was pointed out that
Jordan-Thiry  theory  tries to unify gravity and electromagnetism in a 5
dimensional
space-time. However, in order to reach that goal, to be
self-consistent and to be fully covariant such a theory cannot contain 
(when reduced to 4 dimensions) only these two fields but needs the 
presence of a third field which is a scalar one.

This field is coupled to the other two fields through the field
equations which
arise in a 4 dimensional space-time from the dimensional reduction of the 5
dimensional Einstein equations.  In particular, these equations describe the
induction of the gradient of such scalar field by an electromagnetic field in
much the same way as in electromagetism a magnetic field can arise from the
time variation of an electric field.

On the other hand, the equation of motion of particles subject to
the unified
field, which is the 5 dimensional geodesic equation, when dimensionally reduced
to 4 dimensions contains force terms due to each of the 3 fields individually.
Assuming that  a neutral extended body has a non zero scalar
charge if it is made of
charged particles at the microscopic level, this was shown to generate an
anti-gravity effect in  neutral bodies  if the term due to the
scalar field cancels the gravitational term in the equation of motion.

The author also presented a  new solution for the
field equations corresponding to the propagation of  coupled
scalar-electromagnetic waves with a dispersion relation showing possible
sub and supraluminal propagation depending on the unperturbed
electromagnetic fields present in the region.

The author  conjectured that this effect might be a possible
explanation for the Nodlan-Ralston effect.

The paper by 
Yu.~S.~Vladimirov and S.~I.~Mamontov
"A 6-dimensional geometric model of gravi-electroweak interactions"
states that 
in the framework of a 6-dimensional geometric model of physical
interactions of Kaluza-Klein type it is possible to entirely unify
general relativity with the Weinberg-Salam theory of electroweak
interactions.

In this 6-dimensional model, \\(1) for additional
coordinates the topology of a 2-torus is used, \\(2) the $x^5$- and
$x^6$-dependence of the mixed components of the 6-demensional metric
is introduced and \\(3) the reduction to a 4-dimensional theory is
carried out using the dyad method in a gauge like the 4-dimensional
chronometric one.

It is shown that (a) the hypercharge $Y$ and the
isospin projection $T_3$ are eigenvalues of dyadic operators along
the directions of two additional coordinates, (b) an arbitrariness
in indentifying the mixed components of the metric tensor with the four
physical vector fields is revealed and (c) an arbitrariness in the
definition of spinor physical fields (the first generation lepton
doublet) is established.

In the talk of V.D.Dzhunushaliev
"Multidimensional SU(2) wormhole bounded by two null surfaces"
the  multidimensional  gravity  on  principal   bundle   with
structural $SU(2)$  gauge  group  was  considered \cite{dzh1}.
In this case the multidimensional metric has the following components:
4D Einstein's
metric on the base of bundle, the diagonal metric on the fibre of bundle
(fibre is a gauge group = symmetric space), nondiagonal components
that equivalent (in Kaluza - Klein's sense) to gauge
(electromagnetic or Yang - Mills) fields.
The multidimensional metric has the form:
\begin{eqnarray}
ds^{2} = e^{2\nu (r)}dt^{2} & - & r^{2}_{0}e^{2\psi (r)}\sum^{3}_{a=1}
\left (\sigma ^{a} - A^{a}_{\mu }(r)dx^{\mu }\right )^{2} -
\nonumber\\
& & dr^{2} - a^{2}(r)\left (d\theta ^{2} +
\sin ^{2}\theta d\varphi ^2\right ).
\nonumber
\end{eqnarray}
The ``potentials'' $A^{a}_{\mu }$ have the monopole-like form.
The solution of corresponding Einstein's vacuum equation
gives a multidimensional wormhole located between two surfaces
$(r = \pm a_0)$:
\begin{eqnarray}
\nu & = & -3\psi,
\nonumber\\
a^2 & = & a^2_0 + r^2,
\nonumber\\
e^{-{4\over 3}\nu } & = & {q\over 2a_0}
\cos \left (\sqrt {8\over 3}\arctan {r\over a_0}
     \right ),
\nonumber\\
v & = & \sqrt 6{a_{0}\over r_0 q}
\tan \left (\sqrt {8\over 3} \arctan {r\over a_0}
     \right ).
\nonumber
\end{eqnarray}
The metric on surfaces $(r=\pm a_0)$ is not singular but $ds^2=0$. As the
nondiagonal components are similar to a gauge fields (in Kaluza - Klein's
sense) then in this sense this solution is dual to the black hole
in 4D Einstein - Yang - MIlls gravity: 4D black hole has the
stationary area outside the event horizon but multidimensional
wormhole inside the null surface (on which  $ds^2=0$).
Similar solution was received for standard 5D Kaluza - Klein's
theory \cite{dzh2}.

In the talk of Spiros Cotsakis
"Mathematical Problems in Quadratic and Higher--Order Gravity and
Cosmology" it was stated that
although higher order gravity theories have been considered in the past 
as a means of a first approximation to quantum approaches to gravitation 
and also providing  better singularity behavior (singularity avoidance), 
they  have also been critisised on being unphysical, not always having a 
well--posed initial value formulation and generally leading to field 
equations of higher than second order, a feature that usually makes their 
analyses intractable. Several other ways of motivating further 
study of these theories, in particular use these theories as a testing 
ground for checking which {\em properties} (e.g, black hole entropy, 
inflation, isotropization, recollapse etc.) may prove to be fundamental,
were discussed.  
Also the question was discussed of whether dynamical equivalence between 
the two conformally related spacetimes, which result from the conformal 
transformation that maps the space of higher order theories of gravity to 
that of general relativity plus scalar fields with non-trivial self 
interaction potentials, implies  their physical equivalence. He showed 
that at least for manifolds of a warped product form with base a compact 
Riemannian manifold with negative Ricci curvature and fiber with radius 
the scalar field generated by the conformal transformation, the two 
theories cannot be physically equivalent.  He presented detailed 
pertubation results which support the past instability conjecture namely, 
that all homogeneous and isotropic, physically reasonable  cosmological 
solutions of general relativity are past unstable in the framework of 
higher order gravity theories.  The issue of constrained 
(Palatini) variations for an arbitrary higher order lagrangian with 
general matter couplings and general symmetric connection 
was considered. He  proved a 
theorem stating that the field equations  are identical to 
those obtained from the usual Hilbert variation. This has the interesting 
consequence that it directly provides a generalization of the conformal 
equivalence theorem for arbitrary Weyl geometries.  It was shown that the 
quasi-exponential solution of quadratic gravity is an attractor of all 
homogeneous and isotropic solutions of the general $f(R)$ theory. In 
quadratic theories, de Sitter space is an attractor of the Bianchi IX 
universe provided the scalar 3--curvature does not exceed the value of the 
general scalar field potential associated with the conformal 
transformation. This result settles the cosmic no--hair conjecture in the 
general case without assuming particular forms (for instance exponential) 
for the self--interacting potential. The proof relies on rigorous 
estimates of
the possible bounds of a certain function (called the Moss--Sahni function) and a
nontrivial argument that connects the behavior of that function to the spatial
part of the scalar curvature.
A generalization of the Collins--Hawking
theorem for higher order gravity theories was proved.
The set of homogeneous
and anisotropic cosmologies that can approach isotropy at late times is of
measure zero in the space of all spatially homogeneous universe models.
He also provided a list of open problems in the field.

The paper of 
A.M. Baranov and N.M. Bardushko
"Algebraic classification of Kaluza space with stationary electromagnetic
and scalar fields alone" deals with
an algebraic classification of electromagnetic scalar fields without 
sources and gravitational field potentials in the five-dimensional Kaluza 
space \cite{b1}. A general problem of the algebraic classification 
of the Kaluza space is investigated in \cite{b2}.

Five-dimensional metric components are given as
$G_{ij} = \delta_{ij} = diag(1,-1,-1,-1)$; $G_{j5} = A_j$; $G_{55}=-1$ 
$\mu\nu = 0,1,2,3,5$; $i,j = 0,1,2,3$; $A_j$ are  the electromagnetic 
$4$-potentials. A cylindrical condition with respect to $5$-th coordinate 
is valid. The Weyl $4$-tensor is mapped on a $10$-dimensional bivector real 
space with metric $G_{AB}=diag(1,1,1,-1,-1,-1; 1,-1,-1,-1)$ $(A,B = 
1,2,...,10)$. Then the traceless symmetric $10\times10$ real Weyl matrix 
has the block form  $W =\left ( \begin{array}{cc} 0 & F \\ F & 0 
\end{array} \right)$, where a rectangular $4\times6$ matrix $F$ has 
the block structure $F =\left (\begin{array}{cc} E & e \\ B & b 
\end{array} \right)$ with traceless symmetric $3\times3$ matrices $E = 
(E_{i,j})$; $B = (B_{i,j})$ and vector-columns $e = (E_{i,0})$; $b = 
(B_{i,0})$. Here $E_i$ is an electric field strength and $B_i$ is a 
magnetic displacement vector.

The algebraic classification may be made most simply in the stationary case
when $e = b = 0$. Then $\lambda$-matrix $W(\lambda)$ is described as
\begin{equation}
W(\lambda) =\left ( \begin{array}{cccc}
-\lambda\,I & 0 & E & 0 \\ 0 & -\lambda\,I & -B & 0\\
E & -B & -\lambda\,I & 0\\ 0 & 0 & 0 & -\lambda
\end{array}
\right),
\end{equation}
where $I$ is an identity $3\times3$ matrix. Under elementary
transformations matrix $W(\lambda)$ is reduced to the block diagonal
form $W(\lambda) = diag(I;\; \lambda\,I;\;
\lambda\,I(E^2+B^2-\lambda^2\,I);\; \lambda)$.

A minimal annihilating polynomial of such matrix is $p(\lambda) =
\lambda (\lambda^6 + a_5\lambda^5 + ...+a_1\,\lambda +a_0)$ and a "total"
annihilating polynomial of the $\lambda$-matrix $W(\lambda)$ may be written
as $P(\lambda) = \lambda^4\,p(\lambda)$. Finally the normal
canonical form of $\lambda$-matrix
$C(\lambda) = diag( I,\, I,\, \lambda ,\, \lambda\,I(E^2+B^2-\lambda^2\,I))$
is obtained.

An eigenvalue problem is reduced to the solution of the six power algebraic
equation for $\,\lambda\,$ or a cubic equation for $\,\Lambda =
\lambda^2:\; det(E^2+B^2 - \Lambda\,I) = 0.\,$ The reality of Weyl matrix
gives an existence of four analogs of Petrov algebraic types here:
$I,\,I_a,\,D,\,0,\,$ where the matrix of $\,I_a\,$ type has eigenvalues
for example: $\,\lambda_1 = ... = \lambda_8 = 0;\,\lambda_9 = -\lambda_{10}
= \lambda$.

When there is an electric field strength $E_j$ alone (or $B_j$) then
$$W(\lambda) = 
diag(Q(\lambda),\bar{Q}(\lambda),\lambda I,\lambda),$$
where $\,Q(\lambda) = \lambda I +iE \, (or \,\lambda I +iB),\, i^2 =
-1,\,$ the complex conjugate is denoted by the bar. In particular the
electric field of a charged pencil has the $\,I_a\,$ algebraic type, the
Coulomb field has $\,D\,$ type, and the charged plane has $\,0\,$ type.

In the case of  scalar field $\,g_{55} = \varphi \,$ without 
electromagnetic and gravitational fields 
for locally geodesic coordinates
there exists the traceless
symmetric $10\times10$ real Weyl matrix with a matrix block $\,W =\left (
\begin{array}{cc} 0 & 0 \\ 0 & \Phi \end{array} \right)$, where the $4\times4$
matrix $\,\Phi\,$ corresponds to the scalar field, $\,\Phi =
(\varphi_{,i,j}).\,$

A problem of the algebraic classification is reduced to an eigenvalue problem
of the symmetric real traceless Weyl matrix with the characteristic
equation $\,det(\lambda I)\cdot det(\lambda I)\cdot det(\Phi -
\lambda \Delta) = 0;\,I\,$ is an identity $3\times 3$ matrix. $\,\Delta =
diag(1,-1,-1,-1).\,$ The traceless of Weyl matrix is an equivalent to a wave
equation $\Box{\varphi} = 0$, i.e. we have a massless scalar field. For
example, when $\,\varphi = \varphi(u),\, u = x^0 -x^1\,$ (a plane wave)
we have $\,[(1,1,2)]\,$ algebraic type, which is also the type of an
asymptotic field of spherical and cylindrical scalar waves.

In a stationary case $\Phi = diag(0,S);\, S =(\varphi_{,a,b});\,
\Delta \varphi = 0;\, a,b = 1,2,3.\,$ For a Coulomb-like static scalar field
we have the algebraic type $\,[(1)(1,1)(1)],\,$ i.e. $\,D$ type.

The paper of 
A.M. Baranov "Kaluza space and magnetic charge"
considers a bivector $\,F_{\mu\nu}\,$ as a generalization of the
Maxwell tensor of electromagnetic field in 
Kaluza five dimensional space-time
(with time-like direction alone) with $\,F_{5i} = -A_{5,i} = C_i;\,
C_0 = A_{5,0} = 0\,$ ($i,j = 0,1,2,3$) and the cylindrical condition with
respect to $x^5$ coordinate. 
A dual rotationed the Kaluza space-time  is defined:
$\,\star\,\;$ (see \cite{b1}) as $\;\,\star\,F_{\mu\nu} =
(1/2)\varepsilon_{\rho\mu\nu\alpha\beta}\,F^{\alpha\beta}\,u^\rho,\,$ where
$\,u^\rho = \delta_0^\rho .$  Then $3$-vector $\,\vec{C}\,$ may be
connected with a magnetic displacement $3$-vector in usual space as $\,\vec{C}
= \star\,\vec{B}.$ The standard dual rotation is defined by $\,\ast\,F_{\mu\nu}
= (1/2)\varepsilon_{\rho\mu\nu\alpha\beta}F^{\alpha\beta}\,\lambda^\rho,\,$
where $\,\lambda^\rho = \delta_5^\rho.\,$ Using repeatedly these operations
we have $\,\star^2 = +1;~\ast^2 = -1.\,$

For stationary magnetic fields without current sources we have $\,rot\vec{B}
= 0.$ The solution of this equation is gradient, $\,\vec{B} =
-\nabla{\varphi},\,$ where $\,\varphi\,$ is a magnetic potential. Such
approach does not correspond to the well-known definition of electromagnetic
field tensor by means of $4$-potential. 
Defining by this way
magnetic field may be connected with the gradient vector field $\,\vec{C},\,$
which is the part of $5$-dimensional bivector $\,F_{\mu\nu},\,$ by the
operation $\,\star: \star\,C_a = -\star\,(\nabla{A_5})_a =
(1/2)\,u^{\rho}\varepsilon_{\rho\,5ab0}F_{b0} = B_a\,$ and conversely,
$\,\star\,B_a = C_a = -(\nabla{A_5})_a;\, a,b = 1,2,3.\,$ Thus
in $4$-dimensional space-time $\,\vec{B}\,$ and $\,\vec{C}\,$ are different
vectors: one has the rotational nature and the other has the gradient nature
($A_5 = \varphi\,$ is a magnetic potential).

The first two equations are the usual Maxwell equations and the last
equation is connected with the 5th component of the 5-current density
vector which is a source of a magnetic potential: $\vec{C} = -\nabla\psi
\equiv -\nabla A_5$ and $\Delta A_5 = \Delta \psi = -4\pi j^5$.

A continuity equation for 5-current in 5-dimensional space-time is
$j^{\mu}_{\;\;,\mu} = j^{0}_{\;\;,0} + div \vec{j} = 0$, where 
the cylindrical condition is used. 
The equation is reduced by the dual rotations
$\;\;\star\;\ast\;\;$ into $\; j^{5}_{\;\;,0} = 0\;$, i.e. $j^5$
does not depend on time. In this case one has  \cite{1}
\begin{equation}
\star\ast j^{0} = j^{5} \;\;\;\;\; \hbox{and} \;\;\;\;\;
\ast\star j^{5} = -j^{0},
\end{equation}
where $\;\star\ast \vec{j} = \ast\star \vec{j} = 0$.

In other word,
in the 5-dimensional space-time 
one may connect electric 
and magnetic charges by the dual rotations $\;\ast\;$ and
$\;\star\;$.  It should be noted that the magnetic charges are not 
observed in 4D space-time.  When the point charges are considered the 
components $j^0$ and $j^5$ may be written as $j^0 = {\bf 
e}\delta(\vec{r})$ and $j^5 = {\bf m}\delta(\vec{r})$, where {\bf e} is an 
electric charge and {\bf m} is a magnetic charge; $\delta(\vec r)$ is  the 
3-space $\delta$-function.  Therefore one may write the  connection 
between these charges \cite{1} 
\begin{equation} \star\ast {\bf e} = {\bf m} 
\;\;\; \hbox{and} \;\;\; \ast\star {\bf m} = -{\bf e}.  \end{equation}
Thus, this approach admits an existence of "double-dual"  pairs of static
magnetic and electric charges.

In the paper of F. Burgbacher, C. L\"ammerzahl and A. Macias\\
"Reasons for the space-time to be four-dimensional"
it was shown that
(contrary to statements found in the literature) 
stable atoms may exist in higher--dimensional space--time.

In showing this, only very few fundamental quantum principles are
used.
Two consequences are drawn from this result. First, it is possible to
determine the dimensionality of space--time from the structure of the
spectra of the hydrogen atoms, and second that the Maxwell equations
in higher dimensions are in general non--local and do not obey Gauss
law.

The equation describing a hydrogen atom is the Schr\"odinger
equation.

This can be based on two reasons:

(i) The form of kinetic energy is independent of the dimension.

(ii) If one makes an axiomatic approach to quantum theory
\cite{AL93}, then one arrives at a modified Dirac equation, the
non--relativistic limit of which  gives a Schr\"odinger equation with
the $d$--dimensional Laplacian as kinetic term.

Consequently, one takes as general ansatz for the Hamilton operator for
the hydrogen atom in $d$ spatial dimensions $
H = - \frac{\hbar^2}{2 m} \Delta_d + V(r)$
where $V(r)$ is a spherically symmetric potential. One can show that
(i) a potential $\sim 1/r$ (that is $\kappa = 1$) of a point charge
in higher dimensions leads to stable atoms in higher dimensions, (ii)
the dimensionality enters the atomic spectra thus making it possible
to infer from spectroscopy the three-dimensionality of space, and
(iii) the Maxwell equations have to be modified in order to allow
solutions of the form $1/r$. (Other reasons for the
four--dimensionality of space--time based on the propagation of
helicity states and of Huygen's principle have been given in 
\cite{LaemmerzahlMacias94}.)

The structure of the Maxwell equations change in higher dimensions.
The equation for the electric potential $\phi$ can be given in terms
of Riesz distributions
$$
\Bigl(\overline{\Delta}_d^{\frac{d-1}{2}}* \phi\Bigr)(x) = \left(4\pi
\right)^{\frac{d-1}{2}}\Gamma\left(\textstyle{\frac{d-1}{2}}\right)
\rho(x)
$$
where $\rho(x)$ is the charge density.  
$\overline{\Delta}_d^{\frac{d-1}{2}}$ is the operator which replaces the 
Laplacian in three dimensions. In general this operator is not a differential 
operator. This operator has the feature that Gauss law is no longer 
valid.

In the talk of Thomas Kl\"osch and Thomas Strobl 
"Complete Classification of 1+1 Gravity Solutions"
a classification of the maximally extended solutions for all 1+1
gravity models (comprising e.g.\ generalized dilaton gravity as well as
models with non-trivial torsion) was presented.
No restrictions are placed on the topology of the arising solutions,
and indeed it was found that for generic models solutions on
non-compact surfaces of arbitrary genus with an arbitrary non-zero number of
holes can be obtained. The classical solution spaces (solutions of the field
equations with fixed topology modulo gauge transformations)
are parametrized explicitly and a geometrical interpretation of the
resulting parameters is provided. This allows also to address various
issues in a Hamiltonian treatment and in canonical quantization of gravity.

\end{document}